# Numerical Method for Modeling Nucleation and Growth of Particles that Prevents Numerical Diffusion

A. Khrabry[1], I.D. Kaganovich[2], S. Raman[3], E. Turkoz[3], D. Graves[1]

[1]Princeton University, Princeton, NJ, USA

[2]Princeton Plasma Physics Laboratory, Princeton, NJ, USA

[3]ExxonMobil Technology and Engineering, Annandale, NJ, USA

**Abstract**

State-of-the-art models for aerosol particle nucleation and growth from a cooling vapor primarily use a nodal method to numerically solve particle growth kinetics. In this method, particles that are smaller than the critical size are omitted from consideration, because they are thermodynamically unfavorable. This omission is based on the assumption that most of the newly formed particles are above the critical size and that the subcritical-size particles are not important to take into account. Due to the nature of the nodal method, it suffers from the numerical diffusion, which can cause an artificial broadening of the cluster size distribution leading to significant overestimation of the number of large-size particles. To address these issues, we propose a more accurate numerical method that explicitly models particles of all sizes, and uses a special numerical scheme that eliminates the numerical diffusion. We extensively compare this novel method to the commonly used nodal solver of the General Dynamics Equation (GDE) for particle growth and demonstrate that it offers GDE solutions with higher accuracy without generating numerical diffusion. Incorporating small subcritical clusters into the solution is crucial for: 1) more precise determination of the entire shape of the particle size distribution function and 2) wider applicability of the model to experimental studies with non-monotonic temperature variations leading to particle evaporation. The computational code implementing this numerical method in Python is available upon request.

1. Introduction

Synthesis of nanoparticles from vapor condensation with tailored properties is important for various practical applications. A prime example is the demand for nanoparticles with a narrow size distribution and controllable mean and mode values in catalytic methane pyrolysis. This process aims to produce hydrogen and valuable co-products like carbon nanotubes making the hydrogen economy more attractive (Diab 2022; Hoecker 2016; Kim 2007, 2014; Okeke 2023; Patzschke 2023). Additional applications for tailored nanoparticles include: lithium ion battery electrodes (Liu 2012; Tanaka 2020; Yan 2021), heat pipes (Alphonse 2023; Nazari 2019), photocatalysis for the synthesis of hydrogen (Ahmad 2015; Liu 2017), etc. For the synthesis of nanoparticles with specific properties, an efficient modeling method is essential, capable of accurately predicting the size distribution resulting from vapor condensation.

Modeling homogeneous condensation of vapor leading to formation and growth of liquid clusters is a challenging task because of a very broad range of sizes of forming liquid particles. The particles formed during nucleation and condensation stages include the smallest aerosol particles which consist of several monomers (monomers are simply atoms in case of condensing atomic vapor) to hundreds of atoms that participate in the early stage of cluster nucleation, and particles up to millions to billions of atoms which form during later condensation stages through particle surface growth and particle coagulation (i.e., interparticle collisions and merger into bigger particles).

While in some cases simplified models based on moments of the particle size distribution function can provide all necessary information about the clusters, but, in a general case, full kinetic models are required to predict the entire cluster size distribution and associated properties of the nanoparticles produced (Kim and Kim 2019) and to validate the model or interpret experimental data (Shigeta 2019, 2021; Wyslouzil and Wölk 2016; Zhang 2021). Examples of the cases where simplified moment models might be sufficient are the following. Fast vapor cooling, e.g., vapor expansion through a Laval nozzle (Zhalehrajabi and Rahmanian 2014), and a spark discharge (Maisser 2015). In these cases, the residence time for both vapor and liquid phases is very short, leading to negligible effects from cluster-cluster collisions or coagulation. As clusters nucleate, vapor condenses on them and, shortly after this, the process quenches, because clusters solidify disabling coagulation. For this regime, models that only use the particle size distribution's moments, such as those in (Bilodeau and Proulx 1996; Frenklach and Harris 1987; Frenklach 2002; Friedlander 1983; Nemchinsky and Shigeta 2012) can describe the process reasonably well. They provide average cluster volume, diameter, and its dispersion. There is also the analytical solution (Tacu 2020) available that provides an explicit expression for these quantities after the nucleation burst. Another example where simplified moment models might be sufficient is the opposite case when the residence time is sufficiently long, and cluster coagulation plays a dominant role (Hoecker 2017). In this case, a self-similar solution (Frenklach 1985; Friedlander and Wang 1966; Friedlander 2000; Lee 1984) describes the shape of the cluster size distribution in normalized variables, and average cluster size can be obtained from even simpler monodisperse model, such as in (Kappler 1979; Kruis 1993; Panda and Pratsinis 1995; Yatom 2018). The intermediate case of moderate cooling rate and residence time, when both nucleation and coagulation affect the final cluster size distribution function, cannot be treated by simplified moment models and requires full modeling of the cluster size distribution. Another important case where full modeling of the clusters size distribution is required is for modeling of cluster growth experiments in flow tubes with a non-monotonic temperature profile such as those employed, for example, in (Hoecker 2017), where clusters form first in a colder gas where temperature is below condensation point but then they evaporate in the regions where the gas temperature becomes higher.

Commonly, a sectional/nodal method (i.e., the Nodal solver of General Dynamics Equation (NGDE; Gelbard 1980; Jacobson and Turco 1995; Mitrakos 2007; Pilinis 2000; Prakash 2003; Zhang 2020) is used to model evolution of the cluster size distribution function. This method lumps clusters into groups (nodes) according to the clusters' size which allows for easier numerical solution. However, this method is prone to numerical diffusion leading to artificial broadening of the cluster size distribution and therefore increasing the number of large-size particles. Also, in this method, the smallest clusters, which are smaller than so-called critical size (i.e., are thermodynamically unstable) consisting of several atoms up to hundreds of atoms, are omitted. The inclusion of these clusters in the model is crucial to model experiments where clusters evaporate due to the temperature increase.

To alleviate the limitations of the NGDE solver, we propose an efficient and accurate low-diffusion numerical method to model the entire clusters size distribution function starting from the smallest clusters consisting of just two atoms. We show in this paper that the proposed scheme is more accurate and even less computationally costly compared to the traditional NGDE approach.

The paper is organized as follows. Section 2 briefly introduces the kinetic equation of cluster growth, or the General Dynamics Equation. Section 3 describes the proposed numerical approach to solving the kinetic equation of cluster growth. Section 4 compares the new method for solving the cluster growth equation to the results of the NGDE solver. And Section 5 provides the summary of results.

## 2. The Kinetic Equation of Cluster Growth - General Dynamics Equation

Evolution of the cluster size distribution is determined by kinetics of cluster formation and growth. It can be described by the general dynamics equation (GDE; Friedlander 2000; Gelbard and Seinfeld 1979) which can be written in a symbolic form:

$$\frac{dn_i}{dt} = \frac{dn_i}{dt}\bigg|_{mono} + \frac{dn_i}{dt}\bigg|_{coag}. \qquad (1)$$

Here $n_i$ represents density of clusters consisting of *i* atoms; the first term in the right-hand side (RHS) describes the density change due to attachment/loss of monomers (atoms) to a cluster; the second term in the RHS describes the density change due to collisions between clusters leading to clusters' coagulation. The first term plays an important role during early stages of condensation when clusters nucleate and grow through surface condensation. The second term comes into play at later stages when the clusters have already formed and most of the vapor has condensed on them, but the clusters continue to grow as they collide with each other and coagulate. This is a much slower process than the cluster growth through surface condensation.

Due to the strong separation of the condensation stages, in multiple models which are primarily concerned with the nucleation and surface growth, the second term is omitted, see e.g. (Friedlander 1983; Girshick and Chiu 1989; Girshick 1990; Tacu 2020). In this paper, we consider a model accounting for both terms, while focusing on a numerical scheme for the first term, because it presented a substantial computational challenge in earlier works.

The first term in the RHS of the cluster size growth Eq. (1) can be defined as:

$$\frac{dn_i}{dt}\bigg|_{mono} = J_i - J_{i+1} = f_{i-1}n_{i-1} - r_i n_i - f_i n_i + r_{i+1}n_{i+1}, \qquad (2)$$

where $J_i$ is a net rate of formation of clusters of size *i* (clusters containing *i* atoms) from clusters of size *i*-1 which, in turn, can be determined from the following kinetic equation:

$$J_i = f_{i-1}n_{i-1} - r_i n_i. \qquad (3)$$

Here, the first term in the RHS is responsible for the forward (growth) rate of the *i*-th cluster formation, due to atom attachment/condensation on the *i*-1-th cluster. The second term in the RHS is responsible for the reverse rate due to evaporation/detachment of monomers from the *i*-th cluster. Forward and reverse rate coefficients $f_i$ and $r_i$ in the RHS of Eq. (3) are defined and described in detail in Appendix 1.

Eq. (1) with corresponding definitions (2-3) of the terms in the RHS describes the kinetics of cluster growth for each cluster size starting from dimers and ending with the largest clusters consisting of billions of atoms. It would be computationally prohibitive to solve this equation directly for each cluster size *i* within this range. A sectional/nodal method (Gelbard 1980; Mitrakos 2007; Prakash 2003) is commonly used to reduce the numerical complexity. In this method, the cluster size space is split in exponentially increasing sections (or nodes) where each node contains clusters having sizes within a given range. The kinetic equation (1) is reformulated for average densities of clusters within each node as follows. We define a node *p* of clusters in the range $[i_p, i_{p+1} - 1]$, where $i_p$ is the smallest cluster size belonging to the node *p*, and $i_{p+1}$ is the smallest cluster size belonging to the next node *p+1*. Average density of clusters in the node *p* is:

$$\tilde{n}_p = \sum_{i_p}^{i_{p+1}-1} n_i / (i_{p+1} - i_p).$$

Summing up Eqs. (1) for cluster sizes within a node *p* and dividing by $(i_{p+1} - i_p)$ yields an equation for node densities $\tilde{n}_p$:

$$\frac{d\tilde{n}_p}{dt} = \frac{d\tilde{n}_p}{dt}\bigg|_{mono} + \frac{d\tilde{n}_p}{dt}\bigg|_{coag}, \tag{4}$$

where, in accordance with Eqs. (2), the majority of the terms in the RHS describing monomer addition and evaporation in subsequent cluster sizes cancel each other out yielding the following expression for the first term in the RHS of Eq (4):

$$\frac{d\tilde{n}_p}{dt}\bigg|_{mono} = \frac{f_{i_p-1} n_{i_p-1} - r_{i_p} n_{i_p} - f_{i_{p+1}-1} n_{i_{p+1}-1} + r_{i_{p+1}} n_{i_{p+1}}}{i_{p+1} - i_p}. \tag{5}$$

Relations (4) and (5) form a system of algebraic equations which is not closed, because the cluster densities $n_{i_p-1}, n_{i_p}, n_{i_{p+1}-1}, n_{i_{p+1}}$ for the cluster sizes at the boundaries of the node (i.e., $i_p - 1$, $i_p$, $i_{p+1} - 1$, and $i_{p+1}$) need to be approximated through the average nodal densities $\tilde{n}_p$, $\tilde{n}_{p-1}$ and $\tilde{n}_{p+1}$ in order to close this system of equations. Typically, a first-order upwind approximation is used in nodal solvers. In this approximation, densities of clusters at the node boundaries are approximated by an average value from one of the neighbor nodes. The choice of the neighbor node for the approximation (whether it should be node *p-1* or node *p*) is determined by the sign of the net flux $J_{i_p}$ defined in Eq (3). If $J_{i_p}$ is positive, meaning that condensation prevails over evaporation and clusters of node *p* form from clusters of node *p-1* (not vice versa), then the density of the clusters in the node *p*–1 determines variation of the cluster density in the node p. In this case, density from the node *p*–1, i.e., $\tilde{n}_{p-1}$, is used in the approximation. If $J_{i_p}$ is negative, meaning that evaporation prevails over condensation, then $\tilde{n}_p$ is used in the approximation. This approximation scheme can be formalized as follows:

$$n_{i_p-1} = n_{i_p} = \begin{cases} \tilde{n}_{p-1}, & \text{if } f_{i_p} > r_{i_p} \\ \tilde{n}_p, & \text{if } f_{i_p} < r_{i_p} \end{cases}. \tag{6}$$

Values $n_{i_{p+1}-1}$ and $n_{i_{p+1}}$ are defined similarly:

$$n_{i_{p+1}-1} = n_{i_{p+1}} = \begin{cases} \tilde{n}_p, & \text{if } f_{i_{p+1}} > r_{i_{p+1}} \\ \tilde{n}_{p+1}, & \text{if } f_{i_{p+1}} < r_{i_{p+1}} \end{cases}.$$

Values of rate coefficients $f_{i_p-1}, r_{i_p}, f_{i_{p+1}-1}, r_{i_{p+1}}$ can also be calculated at the nodes $i_p$, $i_{p-1}$ and $i_{p+1}$ using the same approach. Substitution of the definition (6) in Eq. (5) results in a numerical scheme almost identical to one used in the NGDE code (Prakash 2003) with a minor variation: in the NGDE code, the difference $i_{p+1} - i_p$ in the denominator in the RHS of Eq. (5) can be substituted with $i_p - i_{p-1}$ based on the same criterion as in Eq. (6). This approximation scheme is known for its numerical stability. However, it is also known as being likely to exhibit a strong numerical diffusion (Tsang and Rao 1988) which is detrimental for the solution accuracy as it artificially "smears" the solution over multiple computational nodes leading to artificial broadening of the size distribution.

Additionally, another simplification is commonly used in the nodal solvers (such as the NGDE code; Prakash 2003) for the smallest clusters. The growth of smallest clusters is not numerically modeled but is described by an analytical solution. It is known from the classical nucleation theory (CNT; Bakhtar 2005; Girshick and Chiu 1990; Girshick 1990; Smirnov 2000, 2010) that the formation of clusters below the so-called critical size (Frenkel 1955; Smirnov 2006) is energetically unfavorable as an energy barrier is created due to a large contribution from cluster surface energy into the Gibbs energy of cluster formation. Accordingly, the number density of sub-critical clusters (i.e., clusters smaller than the critical size) decays exponentially with cluster size. Based on this knowledge, it is assumed that the total number of clusters below the critical size is small, and they can be omitted from the model. Instead, the clusters are considered to be "born" already having the number of atoms corresponding to the critical size. The rate of "birth" of new clusters is referred to as nucleation rate which is determined using an analytical expression (Girshick and Chiu 1990, Girshick 1990) derived using CNT and its modifications. Accordingly, an analytical nucleation term is added to Eq. (1) for the clusters of critical size $i_{cr}$:

$$\frac{dn_{i_{cr}}}{dt} = \left.\frac{dn_{i_{cr}}}{dt}\right|_{nucl} + \left.\frac{dn_{i_{cr}}}{dt}\right|_{mono} + \left.\frac{dn_{i_{cr}}}{dt}\right|_{coag}. \tag{7}$$

However, the accuracy of this has not been thoroughly tested as well as numerical properties of numerical schemes in the NGDE code. In the next section we propose an alternative, more accurate, approach to solving Eq. (1) numerically and compare its performance to the NGDE code.

### 3. Proposed Approach to Solving the Kinetic Equation of Cluster Growth

We propose a different numerical approach to modeling the first term in the RHS of the kinetic equation (1) on a computational grid. In this approach sub-critical clusters are directly resolved, and numerical diffusion is suppressed.

We start by rewriting the terms in the RHS of Eq. (2) in such a way that it can be viewed as a partial differential equation that has a distinct diffusive-type and a convective-type terms:

$$\left.\frac{dn_i}{dt}\right|_{mono} = -\frac{(f_{i+1}-r_{i+1})n_{i+1}-(f_{i-1}-r_{i-1})n_{i-1}}{2} + \frac{(f_{i+1}+r_{i+1})n_{i+1}-2(f_i+r_i)n_i+(f_{i-1}+r_{i-1})n_{i-1}}{2}. \tag{8}$$

Introducing for brevity the notation $f_i + r_i = \alpha_i$ and $f_i - r_i = \gamma_i$, Eq. (8) reads:

$$\left.\frac{dn_i}{dt}\right|_{mono} = -\frac{\gamma_{i+1} n_{i+1} - \gamma_{i-1} n_{i-1}}{2} + \frac{\alpha_{i+1} n_{i+1} - 2\alpha_i n_i + \alpha_{i-1} n_{i-1}}{2}. \tag{9}$$

The RHS of Eq. (9) can be interpreted as a second-order discrete approximation of two differential terms on an integer grid:

$$\left.\frac{\partial n_i}{\partial t}\right|_{mono} = -\frac{\partial(\gamma(i)n(i))}{\partial i} + \frac{1}{2}\frac{\partial^2(\alpha(i)n(i))}{\partial i^2}. \tag{10}$$

In this interpretation, *i* is not considered as a discrete integer index, instead it is a continuous independent variable of which *n(i)* is a continuous function. $\alpha(i)$ and $\gamma(i)$ are also considered continuous functions of *i* determined using the relations (A8). Somewhat similar approaches for writing the cluster kinetics equation in a continuous form were used in (Brock 1979; Inguva 2022; Katoshevski and Seinfeld 1997a, 1997b; Kumar and Ramkrishna 1997; O'Sullivan and Rigopoulos 2022; Smith 2016; Suck and Pratsinis 1988; Tsang and Rao 1988) and in works of V. Slezov (Slezov and Schmelzer 1994, 2002; Slezov 1996). Though, V. Slezov and N. Smith primarily used it to derive approximate analytical expressions for the cluster size and growth rate, not for numerical solutions; and (Inguva 2022; Katoshevski and Seinfeld 1997a, 1997b; Kumar and Ramkrishna 1997; O'Sullivan and Rigopoulos 2022; Tsang and Rao 1988) considered only a simplified first-order term of the cluster growth through evaporation and condensation where the second term in Eq. (10) was neglected. The first term in the RHS of Eq. (10) is a convective term responsible for transport of *n(i)* in the continuous space of cluster sizes *i* with velocity $\gamma(i)$. The second term is a diffusive term responsible for transport of *n(i)* in the continuous space of cluster sizes *i* with a diffusion coefficient $\alpha(i)$.

These differential terms can now be discretized on a computational grid $i_k$:

$$\left.\frac{\partial n(i_k)}{\partial t}\right|_{mono} = -2\frac{\gamma_i(i_{k+1/2})n(i_{k+1/2}) - \gamma_i(i_{k-1/2})n(i_{k-1/2})}{i_{k+1} - i_{k-1}} + 2\frac{\alpha(i_{k+1})n(i_{k+1}) - 2\alpha(i_k)n(i_k) + \alpha(i_{k-1})n(i_{k-1})}{(i_{k+1} - i_{k-1})^2} \tag{11}$$

Importantly, the steps of this new grid $\Delta i = i_{k+1} - i_k$ do not have to be equal to unity corresponding to an integer index *i*. If the grid steps $i_{k+1} - i_k$ are equal to unity (i.e., if $i_k = k$), then the new discretized equation (11) turns exactly to the original equation (9) if the cluster densities at half-integer index locations $i_{k+1/2}$ and $i_{k-1/2}$ are interpolated linearly from $i_{k-1}$, $i_k$, and $i_{k+1}$. However, with any other grid steps, equation (11) is a discrete approximation of the original discrete equation (9) but on an arbitrary grid. This approach allows increasing the grid steps (reducing the total number of nodes on the grid size) while maintaining solution accuracy due to the second order approximation. First few grid steps can be made equal to unity while subsequent grid steps are increased exponentially. This technique allows obtaining good resolution of the cluster nucleation process at the same time reducing computational burden for larger clusters, similar to how it was done in (Girshick and Chiu 1989).

The major numerical challenge associated with using the approximation (11) originates from the discretization of the convective term in the RHS. The values of the cluster density at half-integer index locations $i_{k+1/2}$ and $i_{k-1/2}$ need to be expressed through values at integer index locations $i_{k-1}$, $i_k$ and $i_{k+1}$. We thoroughly tested performance of several numerical schemes that are commonly used for this purpose (see Appendix 2). Based on the results of these tests, we selected the ULTIMATE-QUICKEST (UQ; Leonard 1991) scheme as the one with the lowest numerical diffusion and good numerical stability.

We applied this scheme in the computations presented in the subsequent section where we compared the performance of our code to the NGDE code.

## 4. Comparing the New Method For Solving The Cluster Growth Equation to the Results of the NGDE Solver

In this section, we present modeling results for condensation of aluminum and iron vapors cooling from initial temperatures of respectively 1500 C and 1800 C and respective equilibrium (saturated) vapor pressures. These initial temperatures are roughly 1300 C below boiling points that would be at atmospheric pressure for both metals and correspond to close initial pressures of 90 Pa and 105 Pa for aluminum and iron vapors respectively. Physical parameters of these materials are summarized in Table 1. The major difference between these metals is manifested in the surface tension coefficient (in the molten state). This difference manifests in different normalized surface energies $\theta$ (as defined in (A5)) which have initial values of 8.1 for Al and 14.5 for Fe. Correspondingly, the energy barrier for nucleation defined in (A9; Frenkel 1955) is considerably higher for Fe than for Al at the same saturation degree. Thereby, though the critical cluster size $i_{cr} = (0.66\ \theta/lnS)^3$ is larger for Fe, total density of Fe sub-critical clusters is expected to be orders of magnitude lower.

*Table 1. Physical parameters of Al and Fe used in the modeling*

| Material | Al | Fe |
|---|---|---|
| Surface tenson $\sigma$, N/m | 0.63 (Rhee 1970) | 1.5 (Ozawa 2011) |
| Wigner-Seitz radius $r_W$, nm | 0.158 | 0.148 |
| Atomic mass, kg | 4.48×10$^{-26}$ | 9.27×10$^{-26}$ |
| Boiling temperature $T_{boil}$, K | 2790 | 3135 |
| Latent heat $L$, kJ/mol | 284 | 349.6 |

Computational grids with close total number of nodes (~100 nodes) were used in both codes (our code and the NGDE code). Necessary modifications have been made to the source code of the NGDE solver to increase the number of nodes (original code only handled 40-node grids). The computational grid used in our code had first 30 computational nodes corresponding to natural numbers of atoms in a cluster (grid steps were equal to unity) to resolve cluster nucleation. Subsequent grid steps grew exponentially with a factor of 1.2. Tests presented in Appendix 2 have shown that such a grid is sufficient to accurately predict evolution of the cluster size distribution. Due to the use of an implicit scheme for time integration, much larger time steps were possible with our model compared to the NGDE code which used an explicit scheme. This made our model by an order of magnitude less computationally expensive (i.e., faster) with a similar computational grid in the cluster size domain.

Two temperature regimes were modeled: 1) cluster formation and growth in monotonically cooling vapor at a constant rate, 2) cluster formation with further evaporation in non-monotonically varying temperature.

### 4.1 *Modeling Condensation of Vapor Cooling Down at a Constant Rate*

Similar to the original tests of the NGDE code (Prakash 2003) and our previous paper (Tacu 2020), here we consider Fe and Al vapors cooling at a rate of $10^5$ K/s. In this section, we present results of two series of computational tests. In the first series of tests, the term responsible for cluster coagulation (the second term in the RHS of Eq. (1)) was not taken into account in both our model and the NGDE solver. This was done to study solely the performance of the numerical schemes for the nucleation and surface growth terms. In the second series of tests, the coagulation terms were included in both models.

Modeling results for the first series of tests are presented in Figs. 1 and 2 for Al and Fe, respectively. Time evolution of the cluster size distribution is shown in Figs. 1(a) and 2(a). The difference between two models is immediately apparent from these figures. Our model considers all clusters, while the NGDE model does not consider sub-critical clusters (i.e., clusters smaller than the critical size). The omittance of sub-critical clusters in the NGDE model affects the entire size distribution. First, the NGDE code does not register any clusters earlier than 2 ms for Al and 4 ms for Fe, when all clusters are sub-critical and have not yet surpassed the nucleation barrier. Our model resolves these clusters. Second, our model shows that densities of sub-critical clusters are orders of magnitude higher than densities of other (super-critical) clusters. The nucleation energy barrier causes an exponential decay of sub-critical cluster densities with the cluster size which ends by an inflection point at the critical size in the cluster size distribution function. After the inflection point, a plateau forms in the size distribution of super-critical clusters. Its width spreads from the critical size to the local maximum of the distribution. For comparison, in the results of the NDGE code, cluster density increases from zero at the critical size toward the maximum (which is a global maximum in the NGDE results, i.e., the mode of the distribution). Whereas in our model this increase is not as drastic, and the size distribution function is overall much flatter. This discrepancy seems to be an artifact of the NGDE model caused by the absence of sub-critical clusters. The deviation between the models is more pronounced for Al vapor than for Fe vapor, for which more clusters are produced in the sub-critical size range.

Additionally, artifacts associated with the numerical diffusion can be observed in the results of the NGDE model. For both materials (Al and Fe), the NGDE code significantly overpredicts densities of larger clusters. The peak of the distribution is lower and shifted towards larger cluster sizes in the results of the NGDE code (as shown by a double-sided arrow). These effects can be explained by high numerical diffusion of the first-order upwind numerical scheme implemented in the NGDE code. This behavior is very similar to the one observed in the results produced by the upwind differencing scheme, as shown in Appendix 2. As a result, average diameter of super-critical clusters is overpredicted by the NGDE code for both materials, as can be observed in Figs. 1(b) and 2(b) for Al and Fe, respectively.

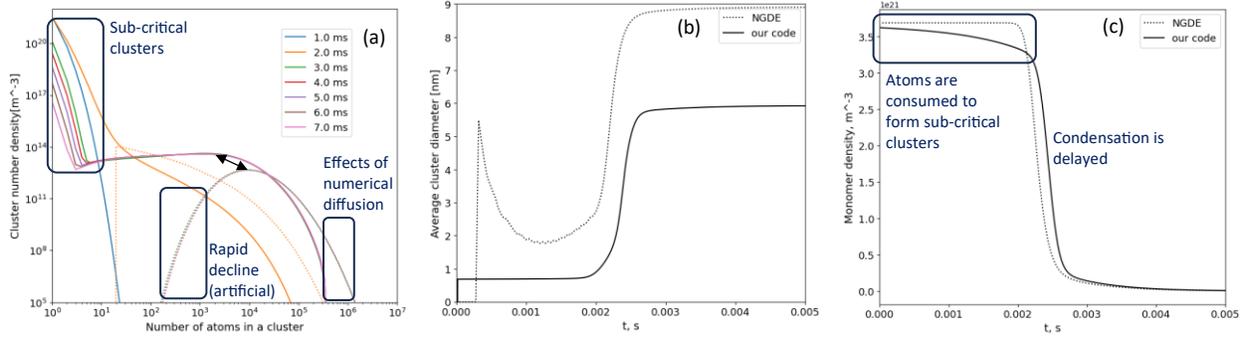

Fig. 1. Comparing modeling results using the new method and the NGDE code for condensation of initially saturated Al vapor cooling down from 1500 C at $10^5$ K/s. Coagulation of the clusters is not modeled. (a) Time evolutions of (a) cluster size distribution, (b) average diameter of the clusters, and (c) vapor atom (monomer) density. Solid lines – results of the new method (i.e., our model), dotted lines – results of the NGDE code. When most of the vapor has condensed, clusters have no material left to grow from, and supercritical cluster size distributions "freeze" after 3 ms.

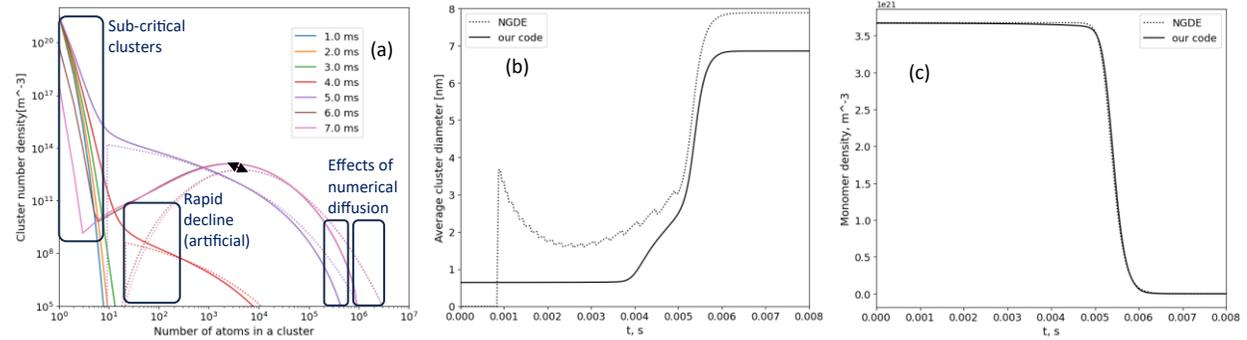

Fig. 2. Modeling results for Fe vapor cooling down from 1800 C at $10^5$ K/s. Coagulation of the clusters is not modeled. Notations are the same as on Fig. 1. When most of the vapor has condensed, clusters have no material left to grow from, and supercritical cluster size distributions virtually "freeze" after 6 ms.

Another implication of accounting for sub-critical clusters in our model is the reduction of vapor density during cooling initial cooling, before rapid condensation happens (see Figs. 1(c) and 2(c)), which occurs because the vapor atoms are consumed by the sub-critical clusters. This effect is considerably stronger for Al which produces sub-critical clusters in greater numbers due to lower surface energy $\theta$, as can be seen in Fig. 1(c). Density of vapor atoms is lower in the results of our model compared to the NGDE code which neglects sub-critical clusters. Lower vapor density implies lower saturation degree $S$ and higher nucleation barrier $4kT\theta^3/(27 ln S^2)$ (Frenkel 1955). As a result, it takes more time for the clusters to surpass the nucleation barrier, and vapor condensation takes longer to occur.

These modeling results are in a good agreement with the analytical solution (Tacu 2020) for the time at which rapid condensation. According to the analytical solution, the relative change in the time of condensation is roughly equal to negative one half of the relative change in the vapor density (given that all other process parameters such as initial temperature and cooling rate remain the same; see Eq. (A16) in Appendix 3). With the sub-critical clusters are accounted, the vapor density is reduced by about 10% (if evaluated at the onset of the rapid vapor condensation). This corresponds to a 5% increase in the condensation time, in good agreement with the numerical solution.

The analytical solution (Tacu 2020) also predicts that the average diameter of super-critical clusters should be proportionally lower when the vapor density is lower (see Eq. (A13) in Appendix 3). This effect contributes to the larger difference in the average cluster size predicted by two model in the case of Al vapor compared to the Fe vapor case. All relevant expressions from the analytical solution (Tacu 2020) are presented in detail in Appendix 3.

The results above have shown that when most of the vapor has condensed, clusters have no material left to grow from, and cluster size distributions virtually "freeze" after 3 ms and 6 ms for Al and Fe respectively. This is an artificial effect due absence of the coagulation term not included in these tests.

Results of the second series of tests where coagulation was included in the model are shown in Figs. 3-6 and discussed below. Longer duration of 100 ms was modeled in the second series of computational tests to allow for the coagulation effects to come into play at later stages of condensation. Similarly to the first series of computational tests, Al and Fe vapors were cooled down during the initial time frame of 8 ms, with the same initial temperatures and cooling rate. After the cooling, the vapors were maintained at constant temperatures. Coagulation terms in our model were implemented in the similar way as it was done in the NGDE code. Integration over all particle sizes was performed to account for inter-particle collisions.

Modeling results for the initial 5 ms of cooling are shown in Figs 3-6(a), 4(c) and 6(c). As evident from these figures, cluster size distributions and average diameter time histories are very similar to the results of previous tests where coagulation was not modeled, with the distinction that the average cluster size keeps growing at a non-zero rate after the nucleation. This result confirms that the effects of coagulation play a minor role during the nucleation stage but become important at a later stage and on longer timescales. Interestingly, the depletion of Al vapor density during the nucleation stage predicted by our model became only stronger when the cluster coagulation was modeled. This is likely the case because the coagulation increases the sub-critical cluster growth rate and thereby accelerates cluster formation and monomer consumption. This effect also accelerates the onset of vapor condensation. This result confirms importance of explicit accounting for the presence of sub-critical clusters in the model.

The difference between the results of two models becomes smaller as the cluster coagulation progresses and initial effects of cluster nucleation fade out. As is evident from Figs. 3(b) and 5(b), cluster size distribution gradually evolves towards a bell-like shape in the results of both models. However, the remnants of the original differences between the models are still observable at later instants: the NGDE code still overestimates densities of larger clusters due to numerical diffusion and underestimates densities of smaller clusters. The difference in the average cluster size predicted by two models is larger of Al clusters (see Figs. 4(b) and 6(b)). Absolute difference in cluster size predicted by two models remains virtually constant for Fe and grows slightly for Al; accordingly relative difference becomes smaller with time as average diameter grows.

Modeling results have been benchmarked by comparing to the self-similar solution derived in (Friedlander and Wang1966; Friedlander 2000) for later stages of cluster coagulation when the effects of the initial shape of the cluster size distribution are diminished. This self-similar solution was derived for normalized variables: normalized number density of clusters is presented as a function of normalized number of atoms in a cluster. The latter is defined as number of atoms in a cluster divided by average number of atoms in a cluster. The former is defined as density of clusters divided by total density of all clusters. Cluster size distributions in normalized variables are plotted on Figs. 3(c) and 5(c). The self-similar solution derived in (Friedlander 2000, see Eq. (7.76) and Table 7.2 therein) is plotted by a black line. Conveniently, the self-similar solution does not depend on material properties, i.e., it is the same for Al and Fe. Not surprisingly, the numerical solutions are noticeably different from the self-similar solution initially, but they gradually approach it as time progresses. This initial difference from the self-similar solution at 5 ms is larger for Fe because the nucleation stage takes longer in Fe vapor. Noticeably, the solutions produced by our model converge to the self-similar solution much faster than those from the NGDE code. We attribute this difference to more efficient low-diffusion numerical scheme used in our model.

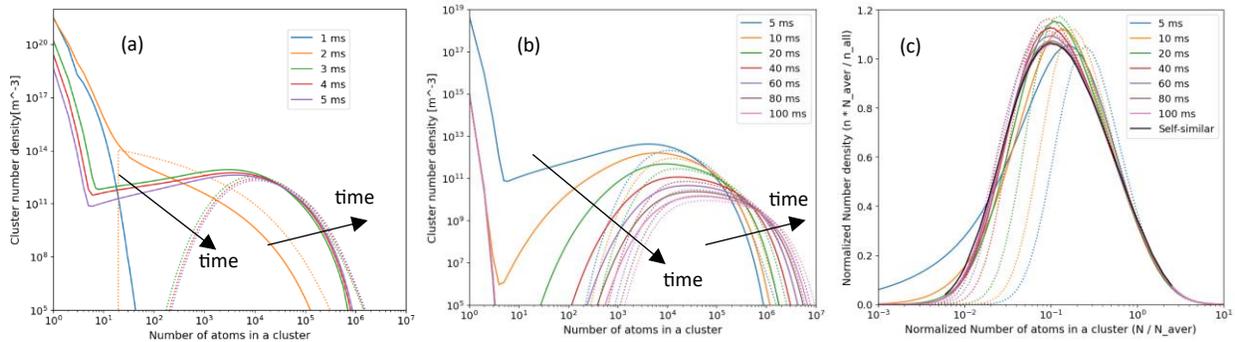

Fig. 3. Comparing modeling results using the new method and the NGDE code for condensation of initially saturated aluminum vapor cooling down from 1500 C to 700 C during the first 8 ms and then remaining at 700 C up to 100 ms. Coagulation of the clusters is modeled. Evolution of the cluster size distribution during the first 5 ms (a), from 5 ms to 100 ms (b) and (c). Solid lines – results of the new method (i.e., our model), dotted lines – results of the NGDE code.

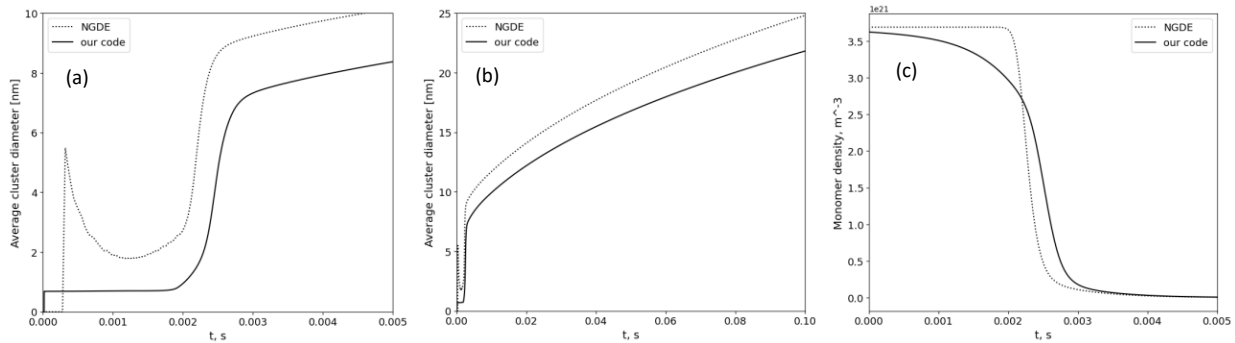

Fig. 4. Comparing modeling results using the new method and the NGDE code for condensation of initially saturated aluminum vapor cooling down from 1500 C to 700 C during the first 8 ms and then remaining at 700 C up to 100 ms. Coagulation of the clusters is modeled. Evolution of the average cluster diameter during the first 5 ms (a) and for the entire residence time (b); evolution of the vapor density (c). Solid lines – results of the new method (i.e., our model), dotted lines – results of the NGDE code.

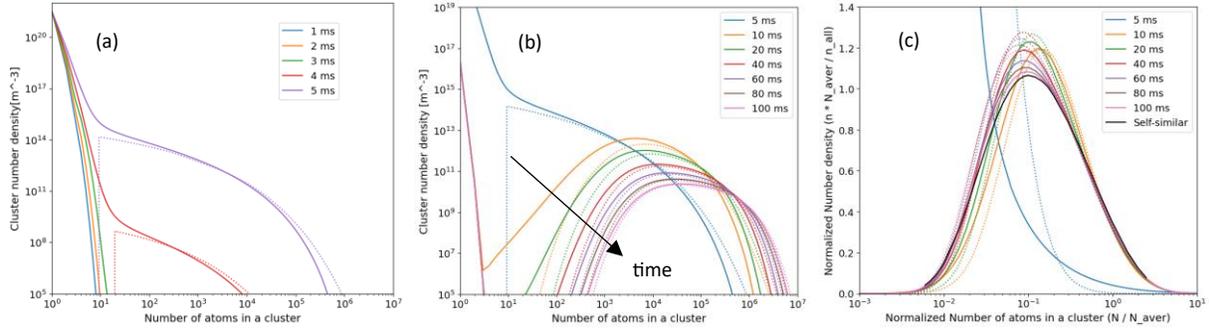

Fig. 5. Modeling results for iron vapor cooling down from 1800 C to 1000 C during the first 8 ms and then remaining at 1000 C up to 100 ms. Notations are the same as in Fig. 3.

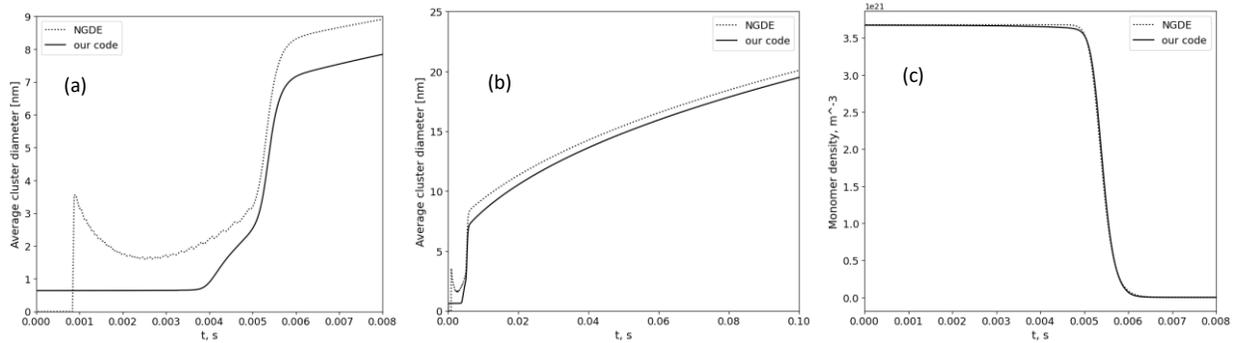

Fig. 6. Modeling results for iron vapor cooling down from 1800 C to 1000 C during the first 8 ms and then remaining at 1000 C up to 100 ms. Notations are the same as in Fig. 4.

## *4.2 Modeling Vapor Condensation and Evaporation in Non-Monotonically Varying Temperature*

In the final series of computational tests, capabilities of the codes to model a condensation process for a non-monotonic temperature profile were compared. Such temperature profiles are used in dedicated experiments on condensation (Hoecker 2016, 2017). The profile was chosen in such a way that vapor condensation followed by evaporation of the clusters. Condensation of initially saturated (S=1) iron vapor was modeled. Temperature profile is shown on Fig 7a: at first, temperature linearly decays for the first 8 ms at the same rate as in the previous tests allowing for the clusters to form. Then the temperature change reverses, and the gas heats back up to the initial temperature of 1800 C during next 8 ms. And finally, the gas is held at this elevated temperature to promote evaporation of the clusters that formed during the first stage. Coagulation of clusters does not play a feasible tole in this process, but it was it was included in the models for consistency. Modeling results for the cooling stage have already been presented in Fig. 5a. Modeling results for the heating and constant temperature stages are plotted on Figs. 7b and 7c respectively. As is evident from these figures, the solutions produced by our model for the heating and constant temperature stages are considerably different from those produced by the NGDE code. Our model predicts evaporation of the clusters during both stages, with cluster densities reducing slightly early in the heating stage (at about 10 ms) and dropping faster as the vapor temperature increases (at about 16 ms). Critical size increases with temperature, and all clusters affectively become sub-critical at the end. At the same time, the NGDE code fails to predict sizeable evaporation of the clusters. The NGDE code was not designed to deal with sub-critical clusters. When the temperature raises back to its initial value and all clusters should become sub-critical, the NGDE model fails to describe this process as it is not included in that model. These results yield an important conclusion that a condensation model needs to explicitly include sub-critical clusters to model condensation processes with non-monotonic temperature profiles featuring both vapor condensation and evaporation of clusters. Explicit inclusion of sub-critical clusters in the model also enables using more accurate Gibbs free energy values for smaller clusters for which the spherical approximation might be inaccurate and so-called 'magic numbers' may come into play (Girshick 2009; Li 2007). We plan to expand on including more accurate thermodynamic data for small clusters in follow-up publications.

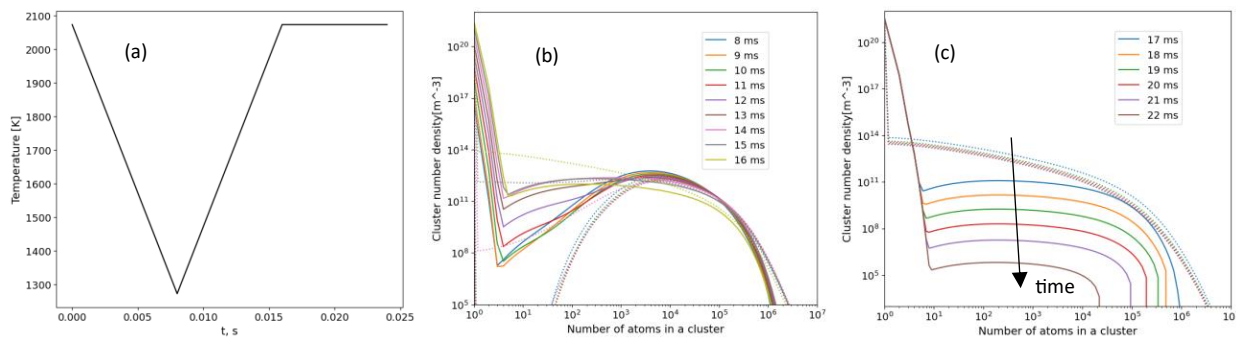

Fig. 7. Modeling results for iron vapor cooling down from 1800 C to 1000 C for the first 8 ms at the same rate as in the previous tests allowing for the clusters to form. During the following 8ms the temperature change reverses, and the gas heats back up to the initial temperature of 1800 C. And finally, the gas is held at this elevated temperature to promote evaporation of the clusters that formed during the first stage. Temperature evolution profile (a), clusters size distributions for time frames from 8 ms to 16 ms (a) and from 17 ms to 22 ms (a). Solid lines – results of the new method (i.e., our model), dotted lines – results of the NGDE code.

## 5. Summary

An accurate numerical method to model nucleation and growth of particles in a condensing vapor has been developed, characterized by low numerical diffusion. It was compared to the conventional nodal method of the NGDE solver. The method proved to be more accurate and less prone to numerical diffusion. The solver developed is an order of magnitude faster than the NGDE code thanks to an implicit time-integration scheme allowing larger time steps. Modeling results have demonstrated that the novel method prevents "smearing" of the cluster size distribution towards larger clusters. Unlike the NGDE code, the new method accounts for sub-critical clusters which has been shown to be important for determining the shape of the cluster size distribution function. In particular, the height of the peak of the cluster size distribution can be underpredicted by the NGDE code by about an order of magnitude and the peak is shifted towards larger cluster sizes due to absence of sub-critical clusters in the model. The inclusion of sub-critical clusters has also been shown to be crucial for the studies of condensation and evaporation processes with non-monotonic temperature evolution where evaporation of clusters is important.

**Acknowledgements**

The authors thank ExxonMobil and Princeton University for funding this project.

**Appendix 1. Definitions of Quantities in the Kinetic Equation of Cluster Growth**

Net rate of formation of clusters *i* from clusters *i-1* is defined in Eq. (3):

$$J_i = f_{i-1} n_{i-1} - r_i n_i$$

Here, $n_i$ represents density of clusters consisting of *i* atoms, $f_i$ and $r_i$ are forward (condensation) and reverse (evaporation) rate coefficients defined as:

$$f_i = \frac{v_{th}}{4} n_1 s_{i-1}, \quad r_i = \frac{v_{th}}{4} s_{i-1} n_1^e \frac{n_{i-1}^e}{n_i^e}. \tag{A1}$$

Here, $v_{th} = \sqrt{8kT/(\pi m_1)}$ is the thermal velocity of the monomers (atoms); $m_1$ is the atom mass; $T$ is temperature; $k$ is the Boltzmann constant; $s_i$ is the surface area of a cluster containing *i* atoms defined as (spherical shape of a cluster is assumed for simplicity):

$$s_i = 4\pi r_W^2 i^{2/3}. \tag{A2}$$

Here, $r_W$ is the Wigner-Seitz radius.

$n_1^e$ and $n_i^e$ are equilibrium densities of monomers and clusters of size *i*, respectively, at a thermodynamical equilibrium at a given temperature. $n_1^e$ can be determined from the Clapeyron-Clausius relation:

$$n_1^e = \frac{1 atm}{kT} \times exp\left(\frac{L}{R}\left(\frac{1}{T_{boil}} - \frac{1}{T}\right)\right),$$

where $L$ is the latent heat of evaporation, and $T_{boil}$ is the boiling temperature at the atmospheric pressure.

The ratio of equilibrium cluster densities in Eq. (A1) can be determined from the detailed balance relation:

$$\frac{n_{i-1}^e}{n_i^e} = exp\left(\frac{\Delta G_i}{kT}\right), \tag{A3}$$

where $\Delta G_i$ is the Gibbs free energy of formation of an *i*-atom cluster from an *i*–1-atom cluster (through an addition of one atom). This Gibbs free energy is associated with the surface energy of the cluster. Most commonly, a simplified model of a spherical cluster is used, in according with the classical nucleation theory (CNT; Bakhtar 2005; Girshick 1990; Girshick and Chiu 1990; Smirnov 2000, 2010). This spherical simplification yields the following Gibbs free energy value in Eq. (A3; Girshick and Chiu 1990):

$$\frac{n_{i-1}^e}{n_i^e} = exp\left((i^{2/3} - (i-1)^{2/3})\theta\right), \tag{A4}$$

where $\theta$ is the normalized surface energy, $\sigma$ is the surface tension of the liquid defined as

$$\theta = 4\pi r_W^2 \sigma/(kT). \tag{A5}$$

The second term in the RHS of Eq. (1) is given by:

$$\left.\frac{dn_i}{dt}\right|_{coag} = \sum_{\substack{j,k \geq 2 \\ j+k=i}} \beta_{j,k} n_j n_k - \sum_{j \geq 2} \beta_{i,j} n_i n_j. \tag{A6}$$

Here, the first term in the RHS describes the rate of formation of clusters of size *i* from collisions of pairs of clusters which have *i* atoms in total. The term in the RHS describes the rate of disappearance of clusters of size *i* through their collisions with clusters of any size. The rates of reverse coagulation processes (i.e., spontaneous splitting of a cluster) are negligible and not included in this equation. $\beta_{i,j}$ is the collision frequency factor as defined in Appendix 1.

Collision frequency factors $\beta_{i,j}$ in the RHS of Eq. (A6) are defined as (Friedlander 2000):

$$\beta_{i,j} = \left(\frac{4\pi}{3}\right)^{5/6} \left(\frac{6kT}{m_1}\right)^{1/2} r_W^3 \left(\frac{1}{i} + \frac{1}{j}\right)^{1/6} \left(i^{1/3} + j^{1/3}\right)^2. \tag{A7}$$

This definition is valid in the case of free molecular collision regime which occurs when the particle size is smaller than collisional mean free path of gas atoms. This regime is typical for the most conditions of interest, i.e., nanometer-scale particles forming at atmospheric pressures.

Coefficients $\alpha(i)$ and $\gamma(i)$ in differential equation (10) are defined as follows:

$$\begin{aligned}\alpha(i) &= \pi v_{th} r_W^2 (i-1)^{2/3} \left(n_1 + n_1^e \exp\left((i^{2/3} - (i-1)^{2/3})\theta\right)\right) \\ \gamma(i) &= \pi v_{th} r_W^2 (i-1)^{2/3} \left(n_1 - n_1^e \exp\left((i^{2/3} - (i-1)^{2/3})\theta\right)\right)\end{aligned}. \tag{A8}$$

Energy barrier for nucleation of the clusters is defined as (Frenkel 1955):

$$\Delta G_i = 4kT\theta^3/(27 \ln S^2). \tag{A9}$$

## Appendix 2. Testing Numerical Schemes for Solving the Kinetic Equation of Cluster Growth

Here, we thoroughly test performance of several numerical schemes that are commonly used for approximating half-integer index values in Eq. (11) including the scheme specifically designed to avoid numerical diffusion and preserve boundedness of the solution at the same time (Khrabry 2010; Leonard 1991) based on the the Convection Boundedness Criterion (CBC; Gaskell and Lau 1988). These schemes are defined below. Expressions are given for the index $i_{k-1/2}$; the values at the index $i_{k+1/2}$ are determined similarly.

1) The first-order upwind differencing (UD) scheme:

$$n(i_{k-1/2}) = \begin{cases} n(i_{k-1}), & \text{if } f_{i_{k-1/2}} > r_{i_{k-1/2}}, \quad i.e., \quad \gamma(i_{k-1/2}) > 0 \\ n(i_k), & \text{otherwise} \end{cases}$$

This is a numerically stable and easy to implement scheme. Similar scheme is effectively used in the NGDE code. However, the accuracy of the scheme only scale as the first order of the grid step, and it is notorious for introducing strong numerical diffusion to the solution.

2) The second-order upwind differencing scheme.

$$n(i_{k-1/2}) = \begin{cases} 1.5\,n(i_{k-1}) - 0.5\,n(i_{k-2}), & if\ \gamma(i_{k-1/2}) > 0 \\ 1.5\,n(i_k) - 0.5\,n(i_{k+1}), & otherwise \end{cases}$$

This scheme is known to be more accurate than the first one, but may still produce some numerical diffusion and exhibit lower numerical stability.

Here, the formulation of this and subsequent schemes is given assuming uniform grid (constant grid steps) for simplicity. In our tests, however, these schemes were implemented in more universal formulations applicable to varying step grids.

3) The third-order upwind differencing scheme also known as QUICK (Leonard 1979) scheme:

$$n(i_{k-1/2}) = \begin{cases} 0.5\,n(i_k) - 0.75\,n(i_{k-1}) - 0.125\,n(i_{k-2}), & if\ \gamma(i_{k-1/2}) > 0 \\ 0.5\,n(i_{k-1}) - 0.75\,n(i_k) - 0.125\,n(i_{k+1}), & otherwise \end{cases}$$

This scheme is notorious for higher numerical accuracy but lower numerical stability.

4) The second-order central differencing (CD) scheme:

$$n(i_{k-1/2}) = 0.5\bigl(n(i_{k-1}) + n(i_k)\bigr)$$

This scheme is, in essence, linear approximation of grid values. This one is notorious for being an accurate scheme but still producing some amount of numerical diffusion.

5) Also, there is a number of specifically designed low-diffusion numerical schemes. Among them, ULTIMATE-QUICKEST (UQ; Leonard 1991) scheme is known for preserving well the shape of an advected profile, i.e., avoiding both numerical diffusion and excessive gradient sharpening (which is a common drawback of low-diffusion schemes) as well as maintaining numerical stability at the same time (Khrabry 2010). It is essentially a QUICK scheme with additional limiter to establish non-diffusive and stability properties. The scheme formulated in terms of so-called normalized variable which makes the overall formulation more cumbersome, and we will not provide it here.

These five numerical schemes introduced in the previous section have been tested through modeling condensation of pure aluminum vapor using a non-uniform (sparse) computational grid. First 30 steps of the grid were equal to unity, and further steps were exponentially increasing with a factor of 1.2. Modeling results are compared to exact numerical solutions of equation (9) obtained with an integer grid. The results are presented in Figs. A1-A3 in the form of cluster size distributions for consecutive time instants: exact solutions of Eq. (9) are plotted by with solid lines; solutions of the approximate equation (11) obtained on the sparse computational grid using various numerical schemes are presented dotted lines (solution obtained with each numerical scheme is shown on a separate plot and compared to an exact solution). It is important to mention here that the solution convergence in regard to time steps has been verified, i.e., presented solutions are perfectly resolved in the time domain and all effects shown here are only associated with various numerical schemes used for approximation of the derivatives in the cluster size domain. First order implicit scheme was used for the time advancement. Time steps were determined from the condition $CFL_i \leq 0.8$ for all computational grid nodes $i$, where $CFL_i$ was defined as $0.25(S - \alpha(i))s(i)n_1^e v_{th}\Delta t/\Delta i$; here $\Delta t$ and $\Delta i$ are steps in time and cluster size domains respectively. These time steps were sufficiently small to deliver both solution accuracy and numerical stability. Collisions or coagulation of the clusters were not modeled in this case, which is a reasonable simplification for these short-time-scale tests.

In Figs. A1 and A2, the results for a model test case of condensation of pure aluminum vapor under constant volume and temperature conditions with initial supersaturation degree S=10 are presented. Fig. A1 shows results for the initial stage of the process (first 0.04 ms). It can be observed how the condensation process initiates as the clusters "climb up" the nucleation barrier with the cluster density exponentially decreasing with the cluster size. The initial critical cluster size at these conditions is $i_{cr} = \left(\frac{2\theta}{3lnS}\right)^3 = 13$ with the dimensionless surface energy $\theta = 8$. Once the clusters surpass the nucleation barrier (grow beyond 13 atoms), their growth continues unimpededly, "down the energy hill", hence the inflection point on the cluster size distribution chats about $i = 13$. Once the condensation proceeds, the clusters grow in size; the vapor density drops (vapor atoms are shown on the charts as clusters of size 1), the critical size increases accordingly and becomes greater than 100, the energy barrier becomes steeper, the density of smaller clusters decreases, and the cluster size distribution becomes non-monotonic. Qualitatively, this behavior is captured with all numerical schemes used. However, as is clear from Fig. A1, numerical schemes UD and CD "smear" the cluster size distribution towards large cluster sizes due to numerical diffusion. The exact solution shows a sharp drop of the cluster density at the cluster size ~$10^4$, however, these numerical solutions predict a smooth density reduction and thereby overpredict density of the clusters above $10^4$ in size. While the second order scheme CD captures density of smaller clusters accurately, the first order scheme UD leads to a substantial distortion in the entire cluster size distribution. The schemes SOUD and QUICK appear to be unstable for this modeling and result in non-physical fluctuations in the cluster density profile. The scheme UQ appears to be the most accurate in predicting the evolution of the cluster size distribution. In captures the shape of the profile well and does not overpredict densities of large clusters.

Cluster size distributions during later in the condensation process, up to 10 ms, are plotted in Fig. A2. The results for the schemes SOUD and QUICK are not shown as they appeared numerically unstable. Cluster sizes have increase substantially since the earlier stage, however qualitative "behavior" of the numerical schemes expanded to this later stage: similarly to the early condensation stage, the scheme UQ produced results sowing the best agreement with the exact solution. Noteworthy that the exact solution is very computationally expensive in this case as numerical grid of the size on order of $10^7$ needs to be used. The schemes UD and CD "smear" the cluster size distribution towards large clusters with this effect being more pronounced with the UD scheme resulting in slight underprediction of medium-sized cluster densities.

Modeling results for a more realistic case of a rapidly cooling vapor with initial saturation degree equal to unity are plotted in Fig. A3. As in the previous case, constant volume model was used. To reduce computational cost for the exact solution, fast cooling rate of $10^7$ K/s was modeled corresponding to a time frame of 0.07 ms. The results are qualitatively similar to the results of the previous test: the UQ scheme demonstrated better performance than the other schemes which overpredicted densities of larger clusters. Overall, based on the results of these tests, it can be summarized that numerical schemes UQ and CD are applicable to solving the approximate cluster growth equation (11): they are numerically stable and produce results that are close to the exact solution in the most part of the cluster size range. However, the CD scheme exhibits some amount of numerical diffusion leading to "smearing" of the cluster size distribution towards larger sizes which makes the UQ scheme the first choice. An exponential increase in the computational grid step of a factor 1.2 is sufficiently large to substantially reduce the grid size and at the same time sufficiently small to alleviate adverse effects on the solution accuracy due to the grid non-uniformity. Based on these results, UQ scheme was used in subsequent modeling with a computational grid having first 30 steps equal to unity and further steps exponentially increasing with a factor of 1.2.

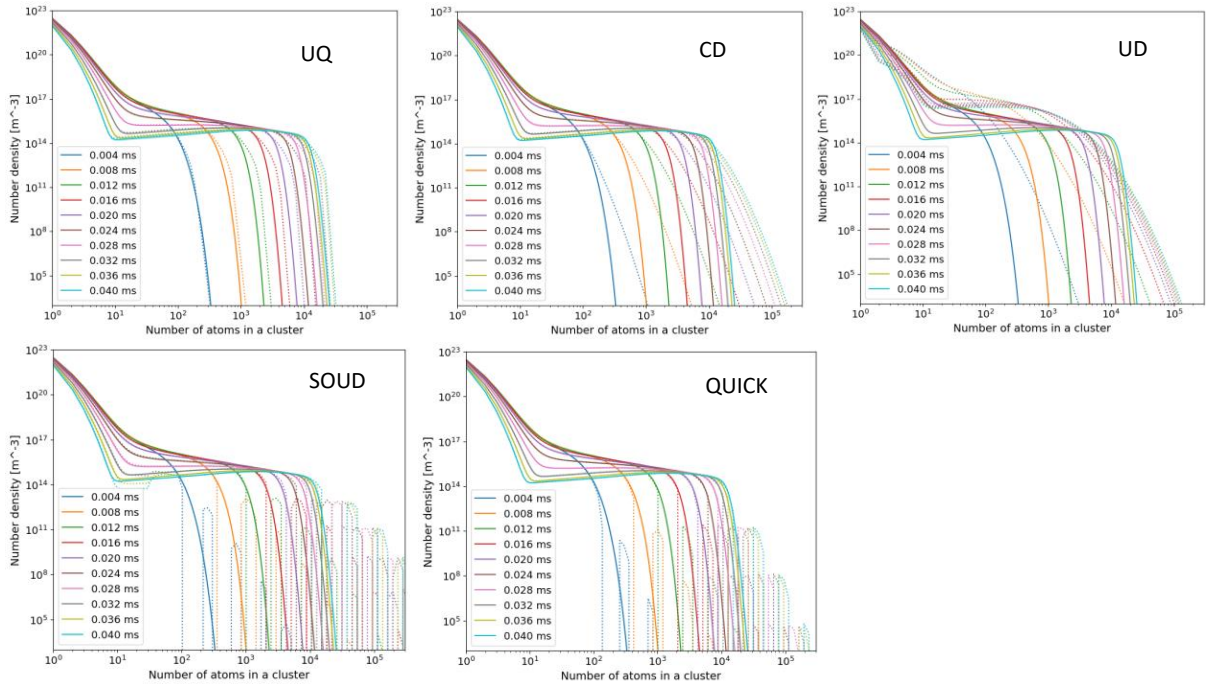

Fig. A1. Time-evolution of the clusters size distribution: modeling results for condensation of initially supersaturated (S=10) aluminum vapor at constant temperature of 1500 C using 5 numerical schemes. Exact numerical solution on a computational grid with a step=1 (solid lines) is compared to numerical solutions obtained using exponentially increasing grid steps. Time frame 0.04 ms.

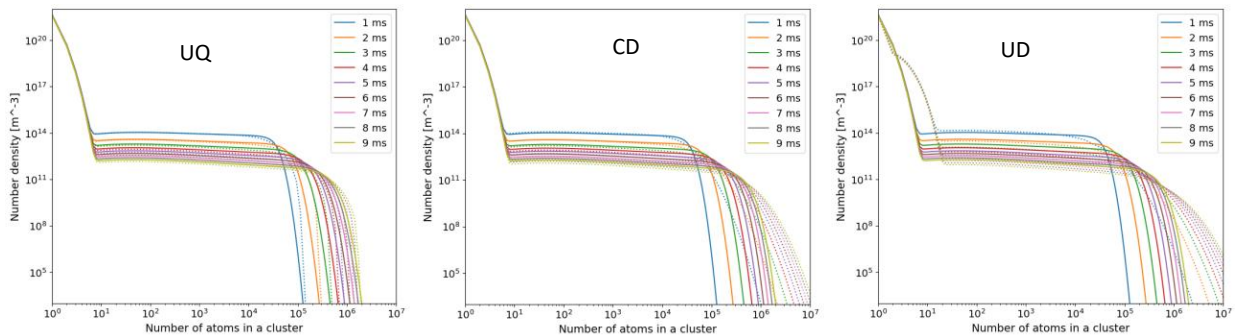

Fig. A2. Time-evolution of the clusters size distribution: modeling results for condensation of initially supersaturated (S=10) aluminum vapor at constant temperature of 1500 C using 3 numerical schemes. Time elapsed is 10 ms.

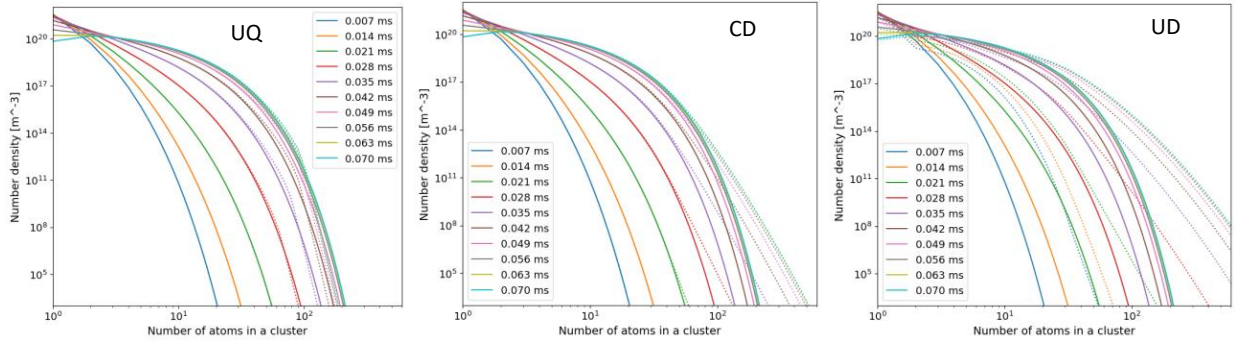

Fig. A3. Time-evolution of the clusters size distribution: modeling results for condensation of initially saturated (S=1) aluminum vapor during rapid cooling from 1500 C with a cooling rate of $10^7$ K/s. Time elapsed 0.070 ms.

## Appendix 3. Analytical Expressions for the Time of Condensation and Average Diameter of Supe-critical clusters from Tacu (2020)

The following analytical expressions from (Tacu 2020) were derived for the vapor is cooling down at a constant rate $\dot{T}_0$ from temperature $T_0$ and initial saturation conditions ($n_1 = n_1^e$ or $S = 1$). According to the solution, the time of the onset of rapid condensation Δt is determined by:

$$\Delta t = \tau_{cooling} \sqrt{\frac{\theta}{27 W(\tau_{cooling}/\tau_{collision})}}. \tag{A10}$$

Here, W(x) is the Lambert W function defined as a solution of the equation:

$$e^W W = x. \tag{A11}$$

$\tau_{cooling}$ and $\tau_{coollision}$ are characteristic time scales of the gas cooling and monomer-cluster collisions defined as following:

$$\tau_{cooling} = \frac{T_0}{\dot{T}_0} \frac{4\pi r_W^2 \sigma N_A}{L}, \quad \tau_{coollision} = \left(\frac{3}{\pi}\right)^{2/3} \frac{4^{1/3}}{v_{th} n_1 r_W^2}. \tag{A12}$$

Here, $N_A$ is the Avogadro number; $v_{th}$, $n_1$ and $\theta$ correspond to the initial conditions (temperature and vapor density).

Mean diameter $d_{mean}$ of particles formed during the rapid condensation is determined by:

$$d_{mean} = 2 r_W \frac{\tau_{cooling}/\tau_{collision}}{W(\tau_{cooling}/\tau_{collision})} \propto \frac{n_1}{\dot{T}_0}. \tag{A13}$$

According to the definitions (A12), in the case of constant cooling rate $\dot{T}_0$, both $\tau_{cooling}$ and $\theta$ are also constant. The only parameter that changes with the monomer density $n_1$ in the RHS of relation (A10) is $\tau_{collision}$. This makes the argument of the function W() proportional to $n_1$. This allows writing a simple relation between the relative of changes of Δt and $n_1$. Differentiation of Eq. (10) yields:

$$\frac{d\Delta t}{\Delta t} = -\frac{1}{2} \frac{dW}{W}. \tag{A14}$$

From the definition of the Lambert W function (A11) it follows that:

$$\frac{dW}{W} = \frac{dx}{x}\frac{W}{W+1},\tag{A15}$$

where $x$ is the argument of the W function, which, we know, is proportional to $n_1$. I.e., $dx/x = dn_1/n_1$. Substituting (A15) into (A14) yields:

$$\frac{d\Delta t}{\Delta t} = -\frac{1}{2}\frac{W}{W+1}\frac{dn_1}{n_1}.\tag{A16}$$

For the conditions considered in this paper the argument of the W function is of order of 1000, which renders W of about 5 and the coefficient W / (W+1) in (A16) close to unity.